# Unusual heavy-mass nearly ferromagnetic state with a surprisingly large Wilson ratio in the double layered ruthenates $(Sr_{1-x}Ca_x)_3Ru_2O_7$

Zhe Qu,[1] Leonard Spinu,[2] Huiqiu Yuan,[3] Vladimir Dobrosavljevic,[4] Wei Bao,[5] Jeffrey W. Lynn,[6] M. Nicklas,[7] Jin Peng,[1] Tijiang Liu,[1] David Fobes,[1] Etienne Flesch,[1] and Z. Q. Mao,[1*]

[1] *Physics Department, Tulane University, New Orleans, Louisiana 70118, USA.*

[2] *AMRI and Physics Department, University of New Orleans, Louisiana 70148, USA.*

[3] *National High Magnetic Field Laboratory, Los Alamos National Laboratory, MS E536, Los Alamos, NM 87545.*

[4] *Department of Physics and NHMFL, Florida State University, Tallahassee, Florida 32306, USA*

[5] *Condensed Matter and Thermal Physics, Los Alamos National Laboratory, Los Alamos, New Mexico 87545, USA.*

[6] *NIST Center for Neutron Research, National Institute of Standards and Technology, Gaithersburg, Maryland 20899, USA*

[7] *Max Planck Institute for Chemical Physics of Solids Noethnitzer Str. 40, 01187 Dresden, Germany.*

**Abstract**

We report an unusual nearly ferromagnetic, heavy-mass state with a surprisingly large Wilson ratio Rw (e.g., Rw ~ 700 for x = 0.2) in double layered ruthenates $(Sr_{1-x}Ca_x)_3Ru_2O_7$ with 0.08 <x< 0.4. This state does not evolve into a long-range ferromagnetically ordered state despite considerably strong ferromagnetic correlations, but freezes into a cluster-spin-glass at low temperatures. In addition, evidence of non-Fermi liquid behavior is observed as the spin freezing temperature of the cluster-spin-glass approaches zero near x ≈ 0.1. We discuss the origin of this unique magneticstate from the Fermi surface information probed by Hall effect measurements.

The Ruddlesden-Popper (RP) series of ruthenates $(Sr,Ca)_{n+1}Ru_nO_{3n+1}$ have attracted widespread attention with their exceptionally rich physical properties, such as unconventional spin-triplet superconductivity[1], itinerant magnetism[2], orbital ordering[3,4], Mott insulator behavior of purely electronic origin[5], and field-tuned electronic nematic phases[6,7]. The close proximity of these exotic states testifies to the delicate balance among the charge, spin, lattice and orbital degrees of freedom in ruthenates, and provides a remarkable opportunity to tune the system to quantum criticality and/or new exotic states using non-thermal parameters such as chemical composition, pressure, and magnetic field.

The double layered ruthenates $(Sr_{1-x}Ca_x)_3Ru_2O_7$ are of particular interest to explore the composition-tuned exotic phenomena because the end members, $Sr_3Ru_2O_7$ and $Ca_3Ru_2O_7$, show distinct properties. $Sr_3Ru_2O_7$ has a paramagnetic (PM) Fermi liquid (FL) ground state[8]. It exhibits a field-tuned metamagnetic quantum critical end-point, near which non-FL behavior and an electronic nematic phase have been observed[6,7]. In contrast, $Ca_3Ru_2O_7$ orders antiferromagnetically at 56 K, and then experiences a metal-insulator transition (MIT) at 48 K[9,10]; giant magnetoresistance was observed and attributed to a bulk spin-valve effect[11–13]. The $(Sr_{1-x}Ca_x)_3Ru_2O_7$ solid solution series has been previously studied using polycrystalline and flux-grown single crystalline samples[14,15]. Those experiments show that the antiferromagnetic (AFM) state of $Ca_3Ru_2O_7$ is suppressed for x< 0.4, and ferromagnetic (FM) order becomes stable even without an external magnetic field[14,15]. However, there exists a remarkable difference between the reported Curie temperatures: $T_c \approx 3K$ for the polycrystalline samples[14] and $T_c \approx 100$ K for the single crystalline samples[15]. This discrepancy may originate from an impurity phase since the RP series easily involves the[16–18] intergrowth of members of different *n*.

In this Letter, we report results of magnetization, specific heat, and Hall effect for $(Sr_{1-x}Ca_x)_3Ru_2O_7$ single crystalline samples grown by the floating-zone technique. A magnetic phase diagram, which reveals novel phenomena, has been obtained due to the improved sample quality. In contrast with previous results[14,15], no spontaneous FM order is found for any Ca content. We find an unusual nearly ferromagnetic, heavy-mass state with a extremely large Wilson ratio in the composition region $0.08 < x < 0.4$ (e.g., $R_w \sim 700$ for $x = 0.2$), where the FM phase was previously reported[14,15]. This state does not evolve into a long-range FM ordered state despite considerably strong FM fluctuations, but freezes into a cluster-spin-glass (CSG) phase at low temperatures. In addition, evidence of non-FL behavior is observed when the CSG phase is suppressed near $x \approx 0.1$.

All samples used in this study were screened carefully by x-ray diffraction and a SQUID magnetometry to ensure their highly pure and untwinned. Since SQUID magnetometry has an extremely high sensitivity to ferromagnetic materials, it guarantees that the selected samples do not contain any FM impurity phases such as $(Sr,Ca)_4Ru_3O_{10}$[18]. SQUID mag-netometry is also used for systematic magnetization measurements on selected samples. Specific heat measurements were performed using a thermal relaxation method, and the Hall effect was using a conventional four-probe method with B // c axis (the longitudinal resistivity component was eliminated by reversing the field direction).

The phase diagram of $(Sr_{1-x}Ca_x)_3Ru_2O_7$ from our study is shown in Figure 1. It con-sists of three regions at low temperatures. (I) $0 \leq x < 0.08$, metamagnetic. A reversible itinerant metamagnetic transition occurs at moderate magnetic fields below $T_M$, the peak temperature in M(T)/B (see Fig 2 (a)). $T_M$ and the metamagnetic transition field $B_{MM}$ decrease with Ca substitution for Sr, down to zero when x is increased to $\sim 0.08$. Such itinerant metamagnetism has generally been interpreted as a field-tuned Stoner transition into a highly polarized magnetic state[6]. (II) $0.4 \leq x \leq 1$, antiferromagnetic: The system first orders antiferromagnetically at the Néel temperature TN and then experiences a MIT at $T_{MIT}$. $T_N$ and $T_{MIT}$ can be identified by anomalies in magnetization as shown in Fig. 2(a), and are consistent with previous results[14,15]. The magnetic structures of $Ca_3Ru_2O_7$ were recently determined by neutron scattering measurements[10,13]; the structure below $T_{MIT}$ is shown in the inset to Fig. 1. (III) 0.08 <x< 0.4, unusual nearly ferromagnetic, heavy-mass. We will focus on this region below since it is the novel phase revealed in this study.

As shown in Fig. 2(a), M/B is enhanced drastically with increasing Ca content for 0.08 <x< 0.4 and T< 10 K, suggesting enhanced magnetic fluctuations. We have estimated the Wilson ratio $R_w =(\pi^2 k_B^2 /3\mu_B^2)(\chi/\gamma_e)$ for these samples using the susceptibility χ measured with 5 mT and zero-field-cooling (ZFC) histories and the electronic specific heat. $R_w$ generally measures the FM correlation strength. For x = 0, $R_w$ is ~10 at 2 K, consistent with the previous result[8]; it increases significantly with increasing Ca content. $R_w$ also strongly depends on temperature and tends to diverge with decreasing temperature. It reaches surprisingly large values at low temperatures, e.g., Rw ~ 700 for x = 0.2 at ~ 2.5 K (see the inset of Fig. 2(a) and the Rw contour plot in Fig. 1). This divergent tendency is cut off due to the formation of the CSG phase (see below), resulting in a maximum of Rw around x = 0.2. Such a large Wilson ratio far surpasses that of any nearly ferromagnetic material such as Pd (Rw ≈ 6 -8), Ni3Ga (Rw ≈ 40)[19], and $Ca_{0.5}Sr_{1.5}RuO_4$ (Rw ≈ 40)[20], indicating that the FM fluctuations in the 0.08 <x< 0.4 range are considerably stronger.

Nevertheless, such strong FM fluctuations do not evolve into a long-range FM ordered state; this is evidenced by the Arrott plot of magnetization shown in Fig. 2(b). The hallmark of a long-range FM state is the presence of a spontaneous magnetization, which manifests itself as positive intercept on the $M^2$ axis in an Arrott plot. The Arrott plots for 0 <x<0.4 exhibit no positive $M^2$ intercept, indicating the absence of long range FM order even for the samples with the strongest FM fluctuations. We note that the intercept for the samples with x = 0.2 and 0.3 approximately equals to zero, indicating that the system is extremely close to a long-range FM order.

This unusual nearly FM state freezes into a CSG phase at low temperatures. For all of the samples in the 0.08 <x< 0.4 range, we observed irreversibility in χ(T) measured at 5 mT between ZFC and field cooling (FC) histories, as well as a strong frequency dependence of the ac susceptibility peak occurring at the frozen temperature $T_f$, which are typical signatures of a glassy state. Fig. 2(c) presents the typical data for these observations.

All unusual magnetic properties described above can be well understood by assuming the system is in close proximity to a two-dimensional (2D) ferromagnet with $T_c$ = 0 K. This scenario allows the existence of strong FM fluctuations without forming a long-range ordered state at finite temperatures, since long-range FM order in 2D on a square lattice is not stable according to the Mermin-Wagner theorem[21]. The formation of the CSG is the way the system releases the entropy at T > 0 K when disorder unavoidably exists in the solid solution.

Our observation of magnetic properties for x< 0.4 differs from early results as pointed out above[14,15]. The $T_c \approx$ 3K weak FM phase observed in polycrystalline samples is probably due to the CSG phase being driven to a long-range FM state by the intergrowth of the FM (Sr,Ca)RuO$_3$ which was indeed found in their samples[2,14]. The $T_c \approx$ 100 K FM state seen in flux-grown single crystals is likely due to the intergrowth of the FM (Sr,Ca)$_4$Ru$_3$O$_{10}$ with $T_c$ $\approx$ 100 K[15,18]. We actually observed a similar FM feature in the impure crystals from our initial batches which were found to involve tiny amounts of (Sr,Ca)$_4$Ru$_3$O$_{10}$ intergrowth.

The evolution of magnetic states with Ca content discussed above was also probed by Hall resistivity $\rho_{xy}$, which is summarized in Figure 3. For the samples in the metamagnetic region, $\rho_{xy}$ displays a linear field dependence with a positive slope, consistent with the previous report[22]. For the samples showing unusual nearly FM states, $\rho^{xy}$ is strongly enhanced. An anomalous Hall effect (AHE) occurs and $\rho_{xy}$ follows the empirical expression (see Fig. 3b):

$$\rho x_y = R_0 B + 4\pi R_S M, \qquad (1)$$

which is usually observed in FM metals and semiconductors[23]. When the system becomes AFM for x ≥ 0.4, $\rho_{xy}$ recovers to a linear field dependence, but changes its sign to negative. The linear field dependence in this region can be understood by considering the magnetic structure of the AFM state[13]; since the FM $RuO_2$ bilayers stack antiferromagnetically along the c axis, the anomalous contribution from the magnetization of FM bilayers cancels, leaving only the normal term $R_0B$ in $\rho_{xy}$.

We further investigated the thermodynamic properties of these various magnetic states through specific heat measurements. Fig. 4 displays the electronic specific heat divided by temperature $C_e/T$ of typical samples. Electron correlations are strongly enhanced for the samples with enhanced FM correlations; $\gamma_e$ reaches large values for these samples, e.g., $\gamma_e$ 0.25 J/Ru-mol $K^2$ at 0.3 K for x = 0.1, comparable to those seen in $Ca_{2-x}Sr_xRuO_4$ with x ≈0.5[20], indicating the formation of heavy-mass quasiparticles. Moreover, we observe an upturn anomaly for 0.08 <x< 0.40, in sharp contrast with the FL behavior observed in $Sr_3Ru_2O_7$. For x = 0.1, $C_e/T$ increases logarithmically below ~10 K and this behavior continues even below $T_f$ (≈1 K) where $C_e/T$ shows a kink (see the inset of Fig. 4). This observation, together with the power-law temperature dependence of $\chi$ (data not shown), suggests the presence of non-FL behaviors at low temperatures in the vicinity of x ≈0.1.

We now wish to discuss the possible origin of these magnetic phase transitions. In general, since Ca ions are smaller than Sr ions, Ca substitution for Sr is expected to introduce structure distortions; this would tune the electronic band structure and possibly induce a magnetic phase transition[20]. Our Hall effect measurements suggest that orbital states are tuned as well. As shown in Fig. 3(c) the normal Hall coefficient $R_0$ changes sign from positive to negative near x ≈0.1 when the system switches from the metamagnetic state to the CSG phase, suggesting that the charge carriers change from hole- to electron-like. This change might be associated with an evolution of a multiple-band effect. Band structure calculations indicate the existence of competing magnetic correlations for $Sr_3Ru_2O_7$[24]: dxy orbital gives rise to purely electron-like bands, which are close to a FM Stoner instability, while both electron-and hole-like bands can be derived from $d_{xz}$ and $d_{yz}$ orbitals, and they have nesting features and thus favor AFM fluctuations. Such a picture of orbital-dependent magnetic correlations is supported by recent experiments on $Sr_3Ru_2O_7$[25,26]. Under this circumstance, the natural interpretation for our observation of the sign change of $R_0$, as well as the enhancement of FM correlations, is that the dxy bands become increasingly dominant with Ca/Sr substitution. For x $\geq$0.4, the transition from the nearly FM, heavy-mass to the AFM state might be related to a structural phase transition since early ARPES studies on this material system reveal a remarkable change in electronic structure near x = 0.4[27] .

In summary, we have established the magnetic phase diagram of double layered ruthen-ates $(Sr_{1-x}Ca_x)_3Ru_2O_7$ using high-quality single crystals. The system approaches an unusual limiting magnetic state with heavy-mass quasiparticles for 0.08 <x< 0.4: while FM correla-tions are extremely strong, the system does not form long-range FM order but forms a CSG phase at low temperatures; non-FL behavior occurs as the CSG phase is suppressed. These behaviors suggest that this state is in close proximity to a 2D ferromagnetic instability.

**Acknowledgments**

We thank C. M. Varma, I. Vekhter, A. V. Balatsky, M. J. Case, Z. Islam, and Y. Liu for useful discussions. Work at Tulane is supported by NSF under grant DMR-0645305, DOE under DE-FG02-07ER46358 and the Research Corporation, work at UNO by DARPA under HR0011-07-1-0031, work in Florida by NSF under DMR-0542026, and work at LANL by NSF, DOE and the State of Florida. HQY also acknowledges the supports from I2CAM and the kind hospitality at MPI-CPFS. Identification of commercial equipment is not intended to imply recommendation or endorsement by NIST.

Electronic address: zmao@tulane.edu

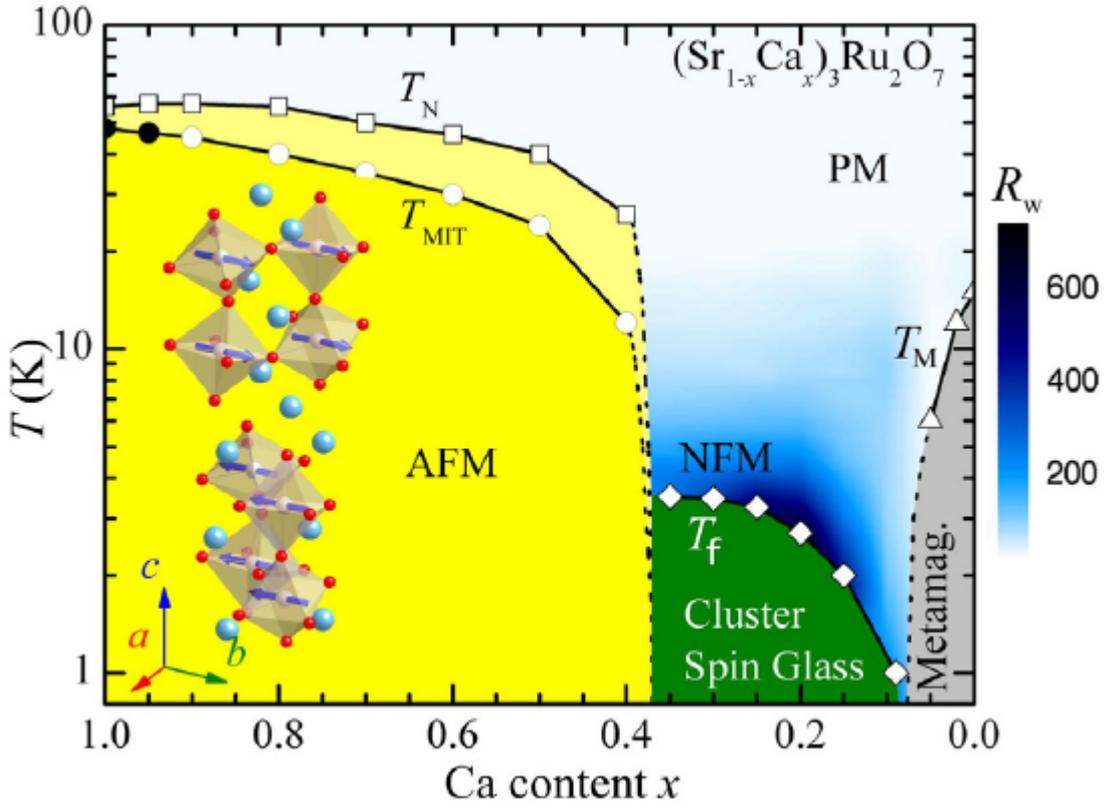

FIG. 1: (Color online) Magnetic phase diagram of $(Sr_{1-x}Ca_x)_3Ru_2O_7$. AFM: antiferromagnetic; $T_N$: the Néel temperature; $T_{MIT}$: the metal-insulator transition temperature. The closed and open circles represent first and second order transition respectively. Metamag.: itinerant metamagnetic; $T_M$: the temperature of the peak in the susceptibility, below which the metamagnetic transition occurs. PM: paramagnetic (blue), represented by the contour plot of the Wilson ratio $R_w$. NFM: unusual nearly FM, heavy-mass state. $T_f$: the frozen temperature of the CSG phase. Inset: the magnetic structure of $Ca_3Ru_2O_7$ below TMIT determined by neutron scattering[13].

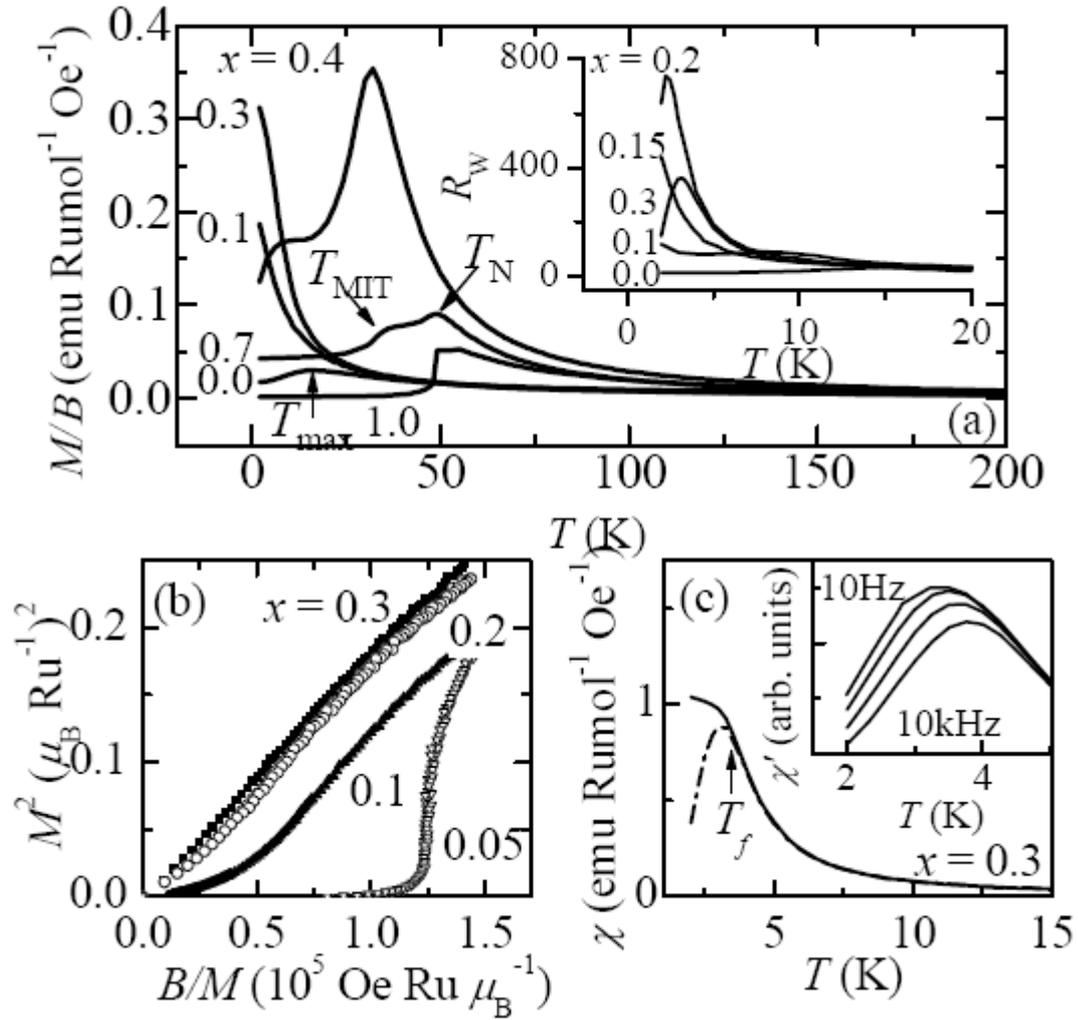

FIG. 2: Magnetic properties for $(Sr_{1-x}Ca_x)_3Ru_2O_7$. (a) Magnetization divided by field M/B vs. T under $B(// a) = 0.3$ T. Inset: the temperature dependence of Wilson ration $R_w$ of typical compositions. (b) The Arrott plot of M(B) at T = 2 K. (c) $\chi(T)$ measured under B = 5 mT with FC (solid line) and ZFC (dotted line) histories for x = 0.3. Inset: ac susceptibility measured at various frequencies for x = 0.3.

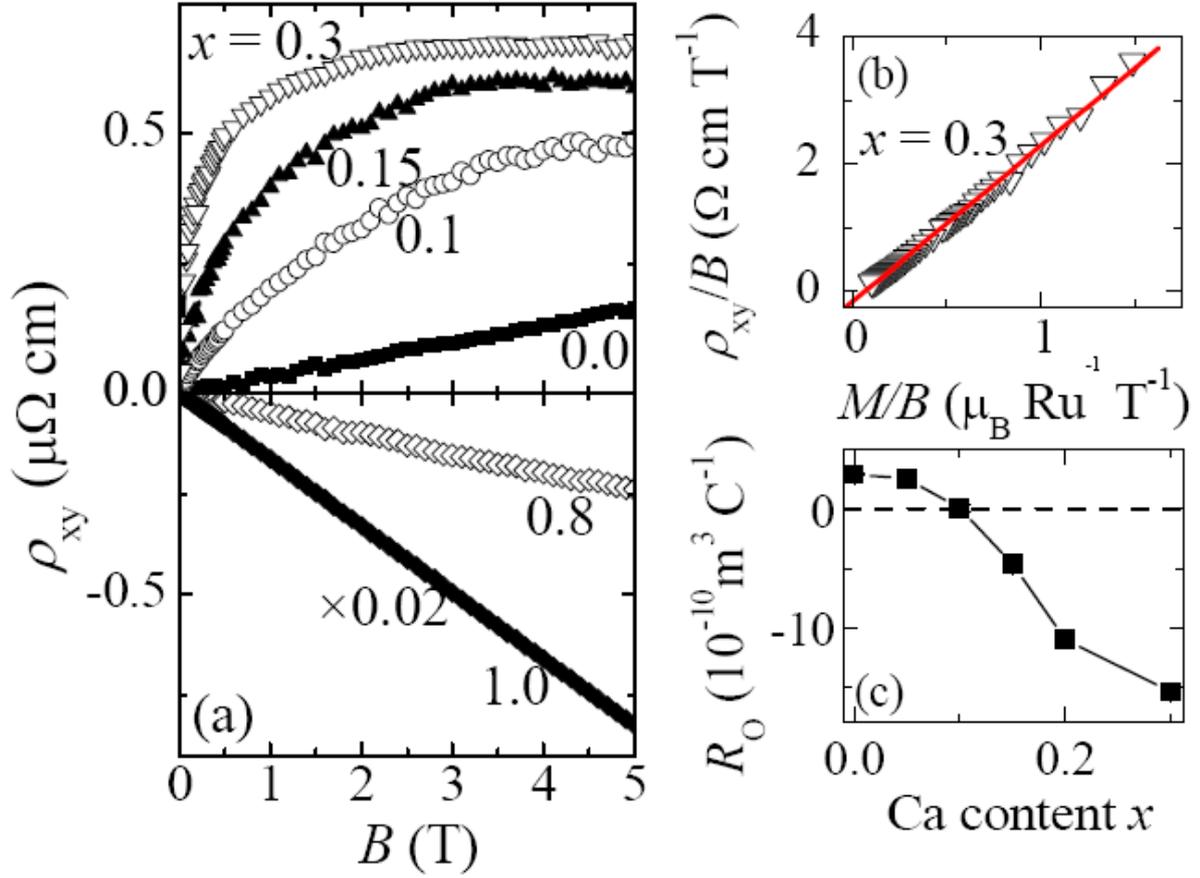

FIG. 3: (Color online) (a) Field dependence of the Hall resistivity $\rho_{xy}(B)$ of various compositions at T = 2 K. (b) $\rho_{xy}/B$ vs. M/B plot for the sample with x = 0.3, which is a typical sample showing AHE. The solid line represents the least-square fitting to Eq. 1. (c) Normal Hall coefficient $R_0$ as a function of x. $R_0$ is determined by fitting the data to Eq. 1 for $0.1 \leq x < 0.4$.

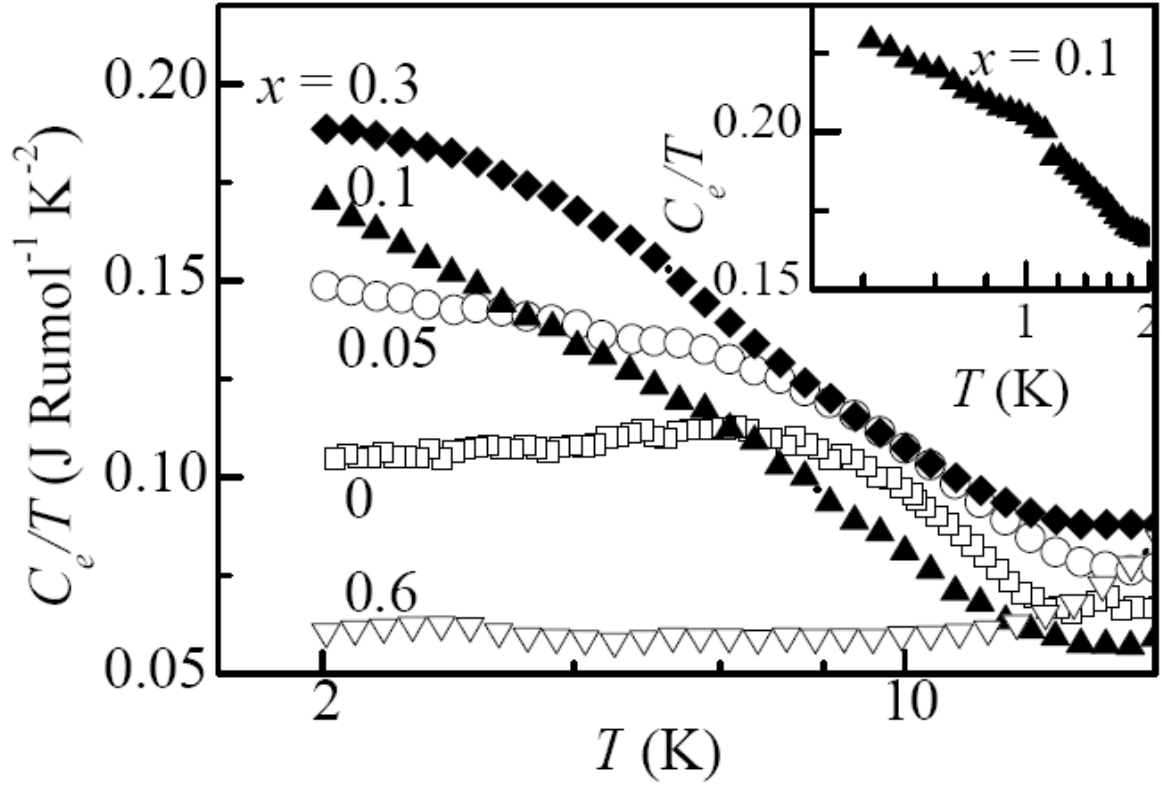

FIG. 4: Electronic specific heat divided by temperature $C_e/T$ vs log$T$ for typical samples of $(Sr_{1-x}Ca_x)_3Ru_2O_7$. Inset: $C_e/T$ vs log$T$ for x = 0.1 down to 0.3 K.

14